\documentclass[aps,prl,reprint,superscriptaddress,amsmath,amssymb]{revtex4-1}

\usepackage{microtype}
\usepackage{xcolor}
\usepackage{graphicx}
\usepackage{dcolumn}
\usepackage{siunitx}
\DeclareSIUnit\Dq{Dq}
\DeclareSIUnit\bohrmagneton{\ensuremath{\mu_{\mathrm{B}}}}
\DeclareSIUnit\wtpercent{\ensuremath{wt.\%}}
\usepackage[colorlinks]{hyperref}

\begin{document}
	
\title{Almost pure $J_{\mathrm{eff}} = 1/2$ Mott state of In$_2$Ir$_2$O$_7$ in the limit of reduced inter-site hopping}

\author{A. Krajewska}
\affiliation{Max Planck Institute for Solid State Research, 70569 Stuttgart, Germany}
\affiliation{Institute for Functional Matter and Quantum Technologies, University of Stuttgart, 70550 Stuttgart, Germany}
\author{T. Takayama}
\affiliation{Max Planck Institute for Solid State Research, 70569 Stuttgart, Germany}
\affiliation{Institute for Functional Matter and Quantum Technologies, University of Stuttgart, 70550 Stuttgart, Germany}
\author{R. Dinnebier}
\affiliation{Max Planck Institute for Solid State Research, 70569 Stuttgart, Germany}
\author{A. Yaresko}
\affiliation{Max Planck Institute for Solid State Research, 70569 Stuttgart, Germany}
\author{K. Ishii}
\affiliation{Synchrotron Radiation Research Center, National Institutes for Quantum and Radiological Science and Technology, Hyogo 679-5148, Japan}
\author{M. Isobe}
\affiliation{Max Planck Institute for Solid State Research, 70569 Stuttgart, Germany}
\author{H. Takagi}
\affiliation{Max Planck Institute for Solid State Research, 70569 Stuttgart, Germany}
\affiliation{Institute for Functional Matter and Quantum Technologies, University of Stuttgart, 70550 Stuttgart, Germany}
\affiliation{Department of Physics, University of Tokyo, Tokyo 113-0033, Japan}

\date{\today}

\begin{abstract}
The pyrochlore iridate In$_2$Ir$_2$O$_7$ is a $J_{\mathrm{eff}} = 1/2$ Mott insulator with frustrated magnetism. Despite the large trigonal distortion, only a small admixture of $J_{\mathrm{eff}} = 3/2$ component in the $J_{\mathrm{eff}} = 1/2$ bands is observed as compared with other pyrochlore iridates A$_2$Ir$_2$O$_7$ (A: trivalent cation). We argue that the reduced inter-site hopping between the $J_{\mathrm{eff}} = 1/2$ and the $J_{\mathrm{eff}} = 3/2$ manifold, due to the large trigonal distortion and the covalent character of In--O bonds, plays a predominant role in the distinct behavior of In$_2$Ir$_2$O$_7$. Such effect of the inter-site hopping should not be dismissed in the local physics of $J_{\mathrm{eff}} = 1/2$ Mott insulators.
\end{abstract}

\pacs{}

\maketitle

%
%

Transition-metal oxides containing $5d$ elements have been receiving attention in the search for novel electronic states generated by the interplay between moderate on-site Coulomb repulsion, $U \sim \SI{2}{\eV}$, and strong spin-orbit coupling, $\lambda_{\mathrm{SOC}} \sim \SI{0.5}{\eV}$~\cite{Witczak-Krempa2014}. Complex iridium oxides have been studied as a platform in this pursuit~\cite{Rau2016}. The common building block in those iridates is an IrO$_6$ octahedron, where five $d$-electrons occupy the $t_{2g}$ manifold. Strong spin-orbit coupling splits the $t_{2g}$ manifold into the completely-filled lower $J_{\mathrm{eff}} = 3/2$ quartet states and the half-occupied upper $J_{\mathrm{eff}} = 1/2$ doublet. $J_{\mathrm{eff}} = 1/2$ states, consisting of the superposition of $d_{xy}$, $d_{yz}$ and $d_{xz}$ orbitals including complex components~\cite{Kim2009}, dominate the transport and magnetic properties.

\begin{figure}[tb]
\centering
\includegraphics[width=0.9\linewidth]{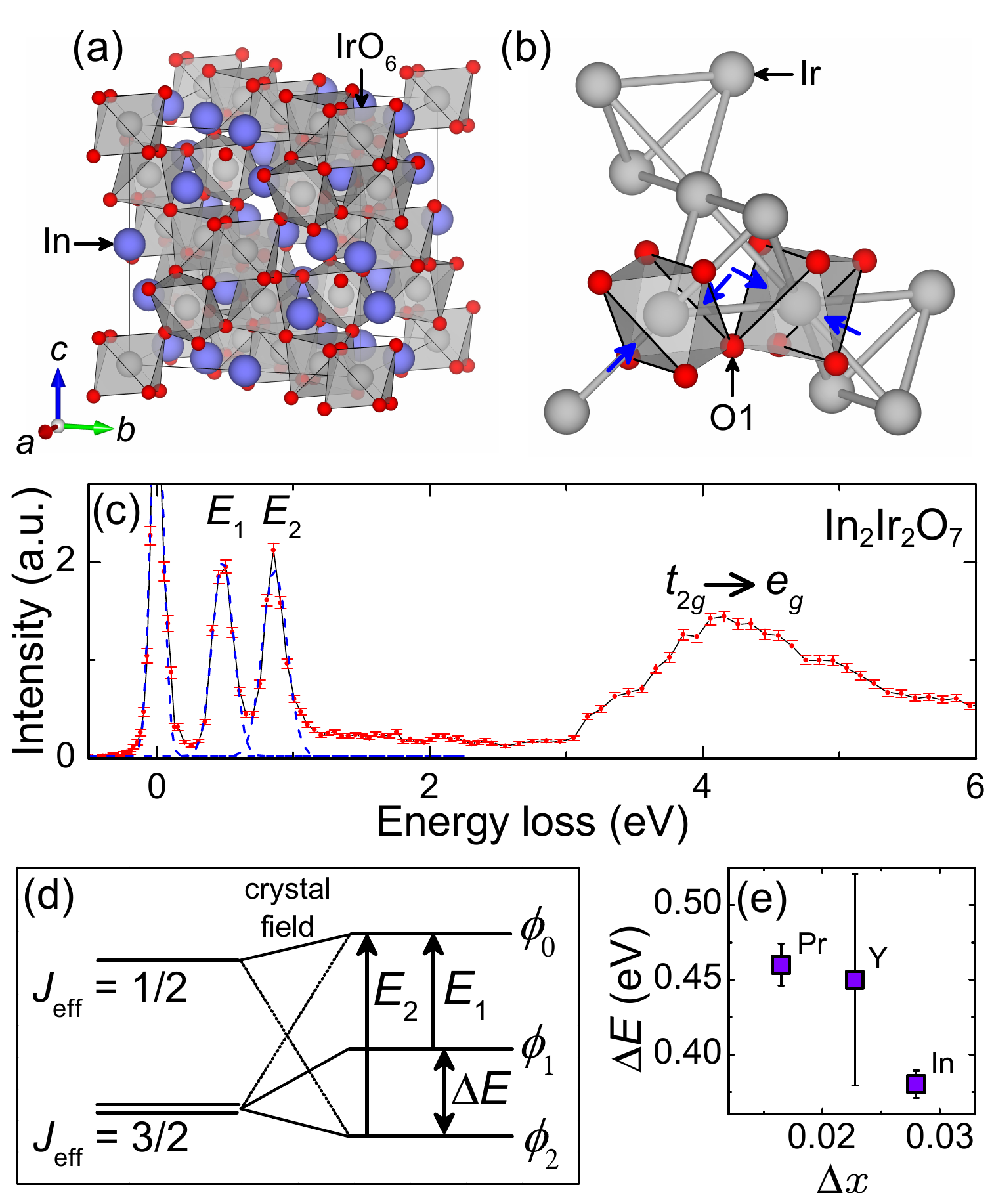}
\caption{(a) Crystal structure of In$_2$Ir$_2$O$_7$. (b) Local structure of two corner-sharing IrO$_6$ octahedra. The blue arrows indicate the direction of trigonal compression. (c) Ir $L_3$-edge RIXS spectrum of In$_2$Ir$_2$O$_7$ polycrystalline sample at $T = \SI{7}{\kelvin}$ with incident energy $E_{i} = \SI{11215}{\electronvolt}$. The Gaussian fits of the peaks are represented by a dashed blue line. The excitations labelled as $E_1$ and $E_2$ correspond to the ones depicted in (d). (d)  Splittings of $J_{\mathrm{eff}} = 1/2$ and 3/2 states into three Kramers doublets ($\phi_0$, $\phi_1$ and $\phi_2$) under trigonal crystal field in a single ion description. (e) The energy split $\Delta{E} = E_2 - E_1$ against $\Delta{x}$ for A$_2$Ir$_2$O$_7$, where A = Pr~\cite{Clancy2016}, Y~\cite{Hozoi2014} and In.}
\label{fig:RIXS}
\end{figure}

Pyrochlore iridates, A$_2$Ir$_2$O$_7$ (A: trivalent ion), where the interplay of $U$ and $\lambda_{\mathrm{SOC}}$ takes place on a frustrated lattice, are expected to host novel electronic phases, including those with non-trivial topological character~\cite{Pesin2010, Wan2011, Witczak-Krempa2012}. A$_2$Ir$_2$O$_7$ crystallize in a cubic structure with space group $Fd\overline{3}m$ consisting of two interpenetrating pyrochlore networks of A and Ir cations (Fig.~\ref{fig:RIXS}(a)). Their electronic ground states evolve with the ionic radius of the A cation~\cite{Matsuhira2007, Matsuhira2011}. The ionic size of A cations in A$_2$Ir$_2$O$_7$ is too small to maintain the regular IrO$_6$ octahedron and therefore the lattice distorts. The IrO$_6$ octahedron is trigonally compressed, which decreases the Ir-O-Ir angle in the neighboring IrO$_6$ octahedra (Fig.~\ref{fig:RIXS}(b)). The bond-bending reduces the inter-site hopping between the neighboring Ir sites and thus controls the $t_{2g}$ bandwidth. The smaller the ionic radius is, the smaller the bandwidth is. With the largest A cation reported, Pr$_2$Ir$_2$O$_7$ is a semimetal with a quadratic band-dispersion and remains metallic down to the lowest temperature~\cite{Kondo2015, Nakatsuji2006}. With decreasing A size, namely reducing the bandwidth, the system undergoes a metal-insulator transition~\cite{Matsuhira2011} accompanied by a magnetic transition. There has been accumulating evidence that an all-in all-out (AIAO) magnetic order is formed~\cite{Sagayama2013, Donnerer2016}. The AIAO ordered phase around the metal-insulator transition is discussed to be a Weyl semimetal~\cite{Ueda2015, Sushkov2015}. With decreasing the size of A cations further to Y$^{3+}$, the system becomes a Mott insulator. While the resistivity shows an insulating behavior up to room temperature, the magnetic transition occurs only at low temperatures.

The trigonal compression of IrO$_6$ octahedra has a pronounced impact also on the local orbital states. The trigonal distortion for an isolated regular IrO$_6$ octahedron splits the $J_{\mathrm{eff}} = 3/2$ quartet into two Kramers doublets~\cite{Rau2014, Uematsu2015}. In addition to the trigonal compression of IrO$_6$, the long-range crystal field, namely that originating from neighboring A and Ir cations, is discussed to be as large as $\sim \SI{0.3}{\eV}$ and thus comparable to or even greater than that from oxygen anions~\cite{Hozoi2014}. The $J_{\mathrm{eff}} = 1/2$-derived state under such trigonal crystal fields should have an admixture of $J_{\mathrm{eff}} = 3/2$ component and cannot be described by the pure $J_{\mathrm{eff}} = 1/2$ wave function~\cite{Liu2012,Winter2017}. In A$_2$Ir$_2$O$_7$ the inter-site hopping between $J_{\mathrm{eff}}$-manifolds may come into play and further modify the $J_{\mathrm{eff}} = 1/2$-derived states~\cite{Propper2016, Sohn2013}. For the elucidation of $J_{\mathrm{eff}} = 1/2$ physics in real materials, consideration of the non-cubic crystal field and the $J_{\mathrm{eff}} = 3/2$ mixing is necessary.

To identify how the $J_{\mathrm{eff}} = 1/2$ state of A$_2$Ir$_2$O$_7$ is modified by the combined crystal field effect from the distortion of IrO$_6$ octahedra and surrounding cations and also by the inter-site hopping, we attempted to synthesize new pyrochlore iridates with a small A cation hosting an even larger trigonal distortion. We discovered a new pyrochlore iridate In$_2$Ir$_2$O$_7$ with non-magnetic In$^{3+}$ ion occupying the A site. In$^{3+}$ ion has the smallest A-site ionic radius (\SI{0.92}{\angstrom}) among the reported A$_2$Ir$_2$O$_7$ pyrochlores and renders the most distorted IrO$_6$ octahedra. In$_2$Ir$_2$O$_7$ is an insulator and shows antiferromagnetic ordering at $T_{\mathrm{m}} = \SI{55}{\kelvin}$. In the resonant inelastic x-ray scattering (RIXS) spectrum, the presence of a sizable splitting of $J_{\mathrm{eff}} = 3/2$ states was identified. Despite the expected enhancement of trigonal crystal field, the magnitude of splitting of $J_{\mathrm{eff}} = 3/2$ states was found to be smaller than those of other A$_2$Ir$_2$O$_7$ with reduced trigonal distortion. Our electronic structure calculations demonstrate that the $J_{\mathrm{eff}} = 1/2$-derived bands of In$_2$Ir$_2$O$_7$ consist predominantly of $J_{\mathrm{eff}} = 1/2$ wave function in contrast to other pyrochlore iridates with comparable or even reduced trigonal distortion. We argue that the long-range trigonal crystal field is responsible for unusual material dependence of the splitting of $J_{\mathrm{eff}} = 3/2$ band and that the inter-site hopping plays a dominant role in bringing $J_{\mathrm{eff}} = 3/2$ character in the $J_{\mathrm{eff}} = 1/2$-derived band.

%
%

The polycrystalline sample of In$_2$Ir$_2$O$_7$ was synthesized by a high-pressure synthesis technique~\cite{SupplementaryData}\nocite{Momma2008, Antonov2004, Shapiro2012, Disseler2012, Yang2017, Clancy2016, Hozoi2014}. The x-ray powder diffraction pattern of the product indicates a high purity phase of pyrochlore In$_2$Ir$_2$O$_7$ with small amounts of impurities ($\sim \SI{3}{\wtpercent}$)~\cite{SupplementaryData}. The Rietveld analysis gives a cubic pyrochlore structure with space group $Fd\overline{3}m$. In$_2$Ir$_2$O$_7$ has the smallest cubic lattice parameter $a = \SI{9.9884(4)}{\angstrom}$ among the reported pyrochlore iridates. The degree of trigonal distortion in cubic pyrochlores can be parameterized by the $x$ fractional coordinate of the O1 atom. $x_{\mathrm{c}} = \num{0.3125}$ corresponds to the case of a regular IrO$_6$ octahedron. $x \neq x_{\mathrm{c}}$ indicates that IrO$_6$ octahedron is trigonally distorted, where $x > x_{\mathrm{c}}$ gives trigonal compression with an Ir-O-Ir angle reduced from the ideal $\sim \SI{141}{\degree}$, commonly found in pyrochlore iridates (Fig.~\ref{fig:RIXS}(b)). $x = \num{0.3405(5)}$ for In$_2$Ir$_2$O$_7$ ~\cite{SupplementaryData} yields the largest trigonal compression and the smallest Ir-O-Ir angle of \SI{125.8(3)}{\degree} among pyrochlore iridates.

{\begin{figure}[tb]
\centering
\includegraphics[width=0.75\linewidth]{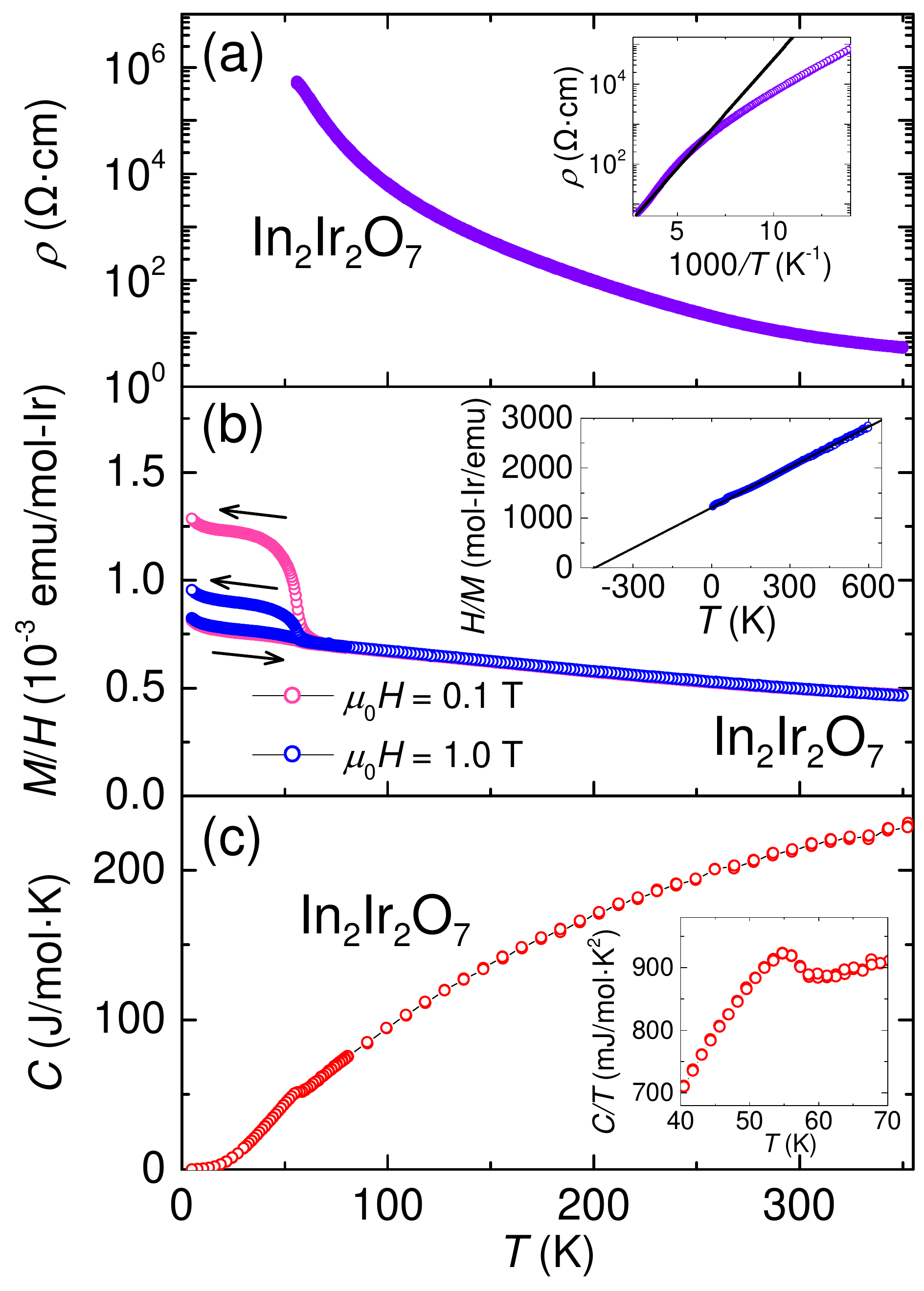}
\caption{Temperature-dependent (a) resistivity $\rho(T)$, (b) magnetic susceptibility $M/H(T)$ and (c) specific heat $C(T)$ of In$_2$Ir$_2$O$_7$ polycrystalline sample. The inset of (a) shows the Arrhenius plot of $\rho(T)$ and the black line indicates the fit at high temperatures. In (b), the black arrows denote the zero-field-cooling ($\rightarrow$) and field-cooling ($\leftarrow$) curves. The inset shows the temperature dependence of inverse magnetic susceptibility, and the black line indicates the Curie-Weiss fit. The inset of (c) shows specific heat divided by temperature around the anomaly at \SI{55}{\kelvin}.}
\label{fig:transport}
\end{figure}}

The resistivity measurement (Fig.~\ref{fig:transport}(a)) shows that In$_2$Ir$_2$O$_7$ is an insulator. From the Arrhenius fit the activation energy around room temperature is estimated to be $\sim \SI{100}{\meV}$ (inset of Fig.~\ref{fig:transport}(a)), which is comparable to those of strongly insulating Ir oxides~\cite{Okabe2011,Singh2012}. Magnetic susceptibility $M/H(T)$ in Fig.~\ref{fig:transport}(b) shows a Curie-Weiss behavior on cooling from room temperature~\cite{MH}. Since In$^{3+}$ is non-magnetic, the magnetic response originates from the 5$d$ electrons of Ir$^{4+}$. The Curie-Weiss fit between \SI{350}{\kelvin} and \SI{600}{\kelvin} yields an antiferromagnetic interaction of $\theta_{\mathrm{CW}} = \SI{-443(7)}{\kelvin}$ and an effective magnetic moment $\mu_{\mathrm{eff}}$ of \SI{1.72(1)}{\bohrmagneton}. The effective moment is close to that of $J_{\mathrm{eff}} = 1/2$ state with $g = \num{2}$, which suggests that the admixture of $J_{\mathrm{eff}} = 3/2$ and other states is not appreciable. At $T_{\mathrm{m}} = \SI{55}{\kelvin}$, an anomaly accompanying a hysteresis between the field-cooling and zero-field-cooling measurements is observed. The history-dependent contribution originates from a weak ferromagnetic moment of 10$^{-4}$ \SI{}{\bohrmagneton}. In the specific heat $C(T)$ (Fig.~\ref{fig:transport}(c)), a clear jump at the onset temperature of the weak ferromagnetic moment is observed evidencing a bulk magnetic transition. Analogous behavior was observed in other pyrochlore iridates such as Y$_2$Ir$_2$O$_7$, where an AIAO magnetic ordering of iridium moments takes place at $T_{\mathrm{m}}$~\cite{Disseler2014}. The small magnetic moment below $T_{\mathrm{m}}$ is discussed to originate from the AIAO domain boundaries~\cite{Hiroi2015}. It is reasonable to ascribe the magnetic transition in In$_2$Ir$_2$O$_7$ to AIAO ordering. The large ratio $f = {\left|{\theta_{\mathrm{CW}}}\right|}/T_{\mathrm{m}} \sim \num{8}$, together with the small magnetic entropy change at $T_{\mathrm{m}}$ ($\sim \SI{5}{\percent}$~of~$R$ln2), implies the presence of strong frustration, as expected from the frustrated pyrochlore lattice.

The Curie-Weiss temperature of $\left|\theta_{\mathrm{CW}}\right| \sim \SI{450}{\kelvin}$ may represent the strength of exchange coupling, which is smaller than those of other insulating pyrochlore iridates such as Y$_2$Ir$_2$O$_7$ ($\sim\SI{1180}{\kelvin}$) and Lu$_2$Ir$_2$O$_7$ ($\sim\SI{725}{\kelvin}$)~\cite{SupplementaryData}. This likely reflects the smaller inter-site hopping $t$ expected in the strongly distorted Ir-O network, since the exchange coupling $J_{ex}$ is scaled by $t^{2}$ as $J_{ex} \sim t^{2}/U$.

\begin{figure}[tb]
\centering
\includegraphics[width=1.0\linewidth]{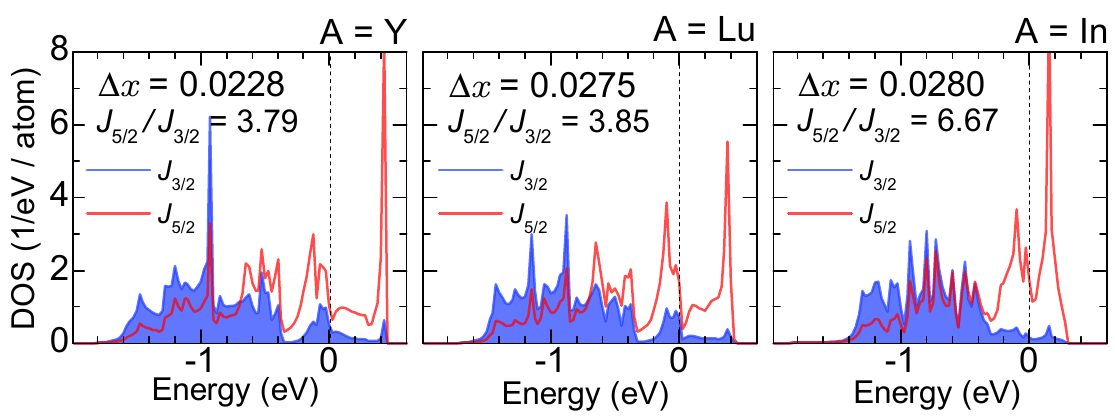}
\caption{$J_{3/2}$ and $J_{5/2}$ resolved DOS for Ir $t_{2g}$ bands in A$_2$Ir$_2$O$_7$ for A = Y, Lu and In. The area beneath the curve of $J_{3/2}$-derived DOS is filled for clarity. The Fermi level $E_{\mathrm{F}}$ is indicated by dashed black lines. The ratio of the area under the $J_{5/2}$- and $J_{3/2}$-derived DOS curves within \SI{\pm 0.5}{\eV} of $E_{\mathrm{F}}$ (\SI{\pm 0.4}{\eV} for A = In) is shown.}
\label{fig:DOS1}
\end{figure}

In order to study the orbital states of In$_2$Ir$_2$O$_7$, we performed RIXS measurement at the Ir $L_3$-edge (Fig.~\ref{fig:RIXS}(c)). In addition to the elastic peak, we observe three clear features at \SIlist{0.48(1);0.86(1);4.14}{\eV}. The two sharp peaks located below \SI{1}{\eV} correspond to excitations within the $t_{2g}$ manifold, while the one at \SI{4.14}{\eV} corresponds to excitations from the $t_{2g}$ to $e_g$ manifold. The split between the $t_{2g}$ and $e_g$ manifold is larger than those of other pyrochlore iridates~\cite{SupplementaryData}. In the absence of distortion in an isolated IrO$_6$ octahedron, only one excitation peak from $J_{\mathrm{eff}} = 3/2$ to $J_{\mathrm{eff}} = 1/2$ would be observed within $t_{2g}$ manifold. The presence of two sharp excitations implies the splitting of $J_{\mathrm{eff}} = 3/2$ quartet into two Kramers doublets ($\phi_1$ and $\phi_2$ in Fig.~\ref{fig:RIXS}(d)), as observed in other pyrochlore iridates~\cite{Uematsu2015,Hozoi2014,Clancy2016}. The widths of two low-energy peaks in the RIXS spectrum of In$_2$Ir$_2$O$_7$ ($\sim \SI{200}{\meV}$) are significantly smaller than those of other reported pyrochlore iridates ($\sim \SI{300}{\meV}$), likely reflecting the smaller bandwidth as identified in the band calculation shown below.

The energy difference of the two low-lying peaks, $\Delta{E} = E_2 - E_1$, represents the splitting of the $J_{\mathrm{eff}} = 3/2$ manifold. We may ascribe the splitting to the trigonal crystal field originating from compression of IrO$_6$. As the strength of this trigonal field can be parameterized by the deviation of the $x$ coordinate of O1 atom from the ideal value ($\Delta{x} = x - x_{\mathrm{c}} = x - \num{0.3125}$), the magnitude of splitting is expected to scale with $\Delta{x}$. In Fig.~\ref{fig:RIXS}(e), $\Delta{E}$ for In$_2$Ir$_2$O$_7$ is compared with those for Y$_2$Ir$_2$O$_7$ and Pr$_2$Ir$_2$O$_7$ with smaller $\Delta{x}$. We find that $\Delta{E}$ does not depend on $\Delta{x}$ appreciably and even decreases with increasing $\Delta{x}$. Factors other than the trigonal distortion of the IrO$_6$ octahedron, such as long-range crystal field from other cations and/or inter-site hopping, should play an important role in the splitting of $J_{\mathrm{eff}} = 3/2$ band.

In order to clarify the combined effect of these factors on Ir $5d$
$t_{2g}$ states, band structure calculations were performed for A$_2$Ir$_2$O$_7$ for A = In, Lu and Y. Y$_2$Ir$_2$O$_7$ has the least trigonally compressed IrO$_6$, while Lu$_2$Ir$_2$O$_7$ and In$_2$Ir$_2$O$_7$ host nearly identical trigonal distortion. Neither Y$^{3+}$ nor Lu$^{3+}$ carries magnetic moments. The density of the Ir $t_{2g}$ states (DOS) projected onto $J = 5/2$ ($J_{5/2}$) and $J = 3/2$ ($J_{3/2}$) states are shown in Fig.~\ref{fig:DOS1}. Since the on-site Coulomb repulsion $U$ is not incorporated, the calculations produce a metallic ground state for all three materials.

The pure $J_{\mathrm{eff}} = 1/2$ wave function comprises only the $J_{5/2}$ states while $J_{\mathrm{eff}} = 3/2$ wave function consists of both $J_{5/2}$ and $J_{3/2}$.  Figure~\ref{fig:DOS1} shows that the states within \SI{\pm 0.4}{\eV} near $E_{\mathrm{F}}$ have $J_{\mathrm{eff}} = 1/2$ character, while those below \SI{- 0.4}{\eV} are $J_{\mathrm{eff}} = 3/2$ ones. Nonetheless, a sizable $J_{3/2}$ weight near $E_{\mathrm{F}}$ is present, indicative of an admixture of the $J_{\mathrm{eff}} = 3/2$ component to the $J_{\mathrm{eff}} = 1/2$ states. Since the trigonal crystal field mixes $J_{\mathrm{eff}} = 1/2$ and $3/2$ states, one would expect more $J_{\mathrm{eff}} = 3/2$, and consequently $J_{3/2}$, weight near $E_{\mathrm{F}}$ in In$_2$Ir$_2$O$_7$ and Lu$_2$Ir$_2$O$_7$ as compared to Y$_2$Ir$_2$O$_7$.  In contrast to this assumption, a comparable contribution of $J_{3/2}$ states appears in Lu$_2$Ir$_2$O$_7$ and Y$_2$Ir$_2$O$_7$, despite the larger distortion in the former. This is evidenced by the ratio of the areas under $J_{5/2}$ and $J_{3/2}$ DOS shown in Fig.~\ref{fig:DOS1}. The $J_{\mathrm{eff}} = 3/2$ weight around $E_{\mathrm{F}}$ is appreciably smaller in In$_2$Ir$_2$O$_7$ than not only that of Y$_2$Ir$_2$O$_7$, but also of Lu$_2$Ir$_2$O$_7$ with a comparable trigonal distortion, distinctly pointing to the formation of almost pure $J_{\mathrm{eff}} = 1/2$ state in In$_2$Ir$_2$O$_7$. The $t_{2g}$ bands of In$_2$Ir$_2$O$_7$ are narrower ($\sim \SI{1.7}{\eV}$) than those of Lu$_2$Ir$_2$O$_7$ ($\sim \SI{2.0}{\eV}$) despite a comparable magnitude of trigonal distortion.

Properties of pyrochlore iridates are expected to vary as a function of the A cation radius, and In$_2$Ir$_2$O$_7$ seemingly adheres to this assumption. Its smallest A cation in the series produces the largest trigonal distortion and the smallest bandwidth, as observed in the band calculation (Fig.~\ref{fig:DOS1}), and manifested in the peak widths of $E_1$ and $E_2$ excitations in RIXS (Fig.~\ref{fig:RIXS}(c)). While the lack of scaling of  $\Delta{E}$ with $\Delta{x}$ might be explained by the contribution of the long-range crystal field, the origin of the contrast of the bandwidth and the orbital character between In$_2$Ir$_2$O$_7$ and Lu$_2$Ir$_2$O$_7$, despite nearly the same degree of trigonal distortion, has yet to be addressed. The inter-site hopping and the covalency of the A--O bonds likely play an important role in fostering the ground state of pyrochlore iridates.

\begin{figure}[tb]
\centering
\includegraphics[width=0.95\linewidth]{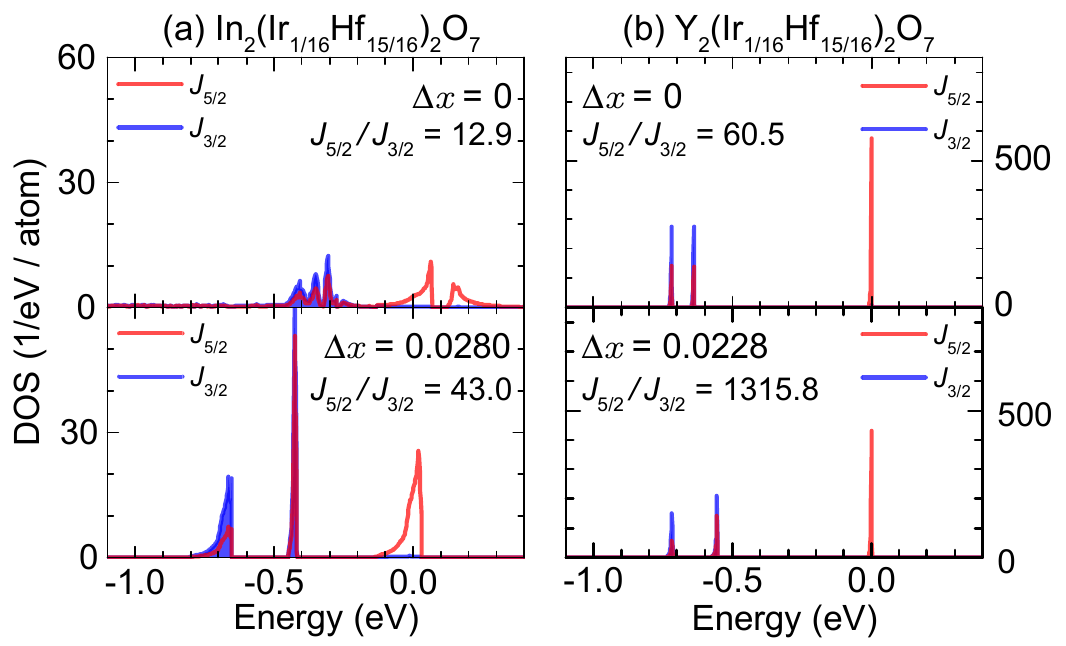}
\caption{$J_{3/2}$ and $J_{5/2}$ resolved DOS for Ir $t_{2g}$ bands for model A$_2$(Ir$_{1/16}$Hf$_{15/16}$)$_2$O$_7$ for A = In and Y for undistorted ($\Delta{x} =0$) IrO$_6$ and the experimentally observed distortion of IrO$_6$. The area beneath the curve of $J_{3/2}$-derived DOS is filled for clarity. The ratio of the area under the $J_{5/2}$- and $J_{3/2}$-derived DOS curves within \SI{\pm 0.3}{\eV} of $E_{\mathrm{F}}$ is shown.}
\label{fig:DOS2}
\end{figure}

To verify our hypothesis, we performed band calculations for model pyrochlore iridates with Ir $d$-$d$ hopping suppressed up to the third neighbors, which is achieved by substitution of 15 out of 16 Ir ions in a cubic unit cell with Hf, giving a chemical formula of A$_2$(Ir$_{1/16}$Hf$_{15/16}$)$_2$O$_7$~\cite{SupplementaryData}. By suppressing hopping and varying the degree of the trigonal compression of the IrO$_6$ octahedron, we attempt to separate the impact of inter-site hopping from that of trigonal crystal fields.

We find that the model compounds with suppressed inter-site hopping (Fig.~\ref{fig:DOS2}) show strongly reduced $J_{\mathrm{eff}} = 3/2$ character near $E_{\mathrm{F}}$ as compared to A$_2$Ir$_2$O$_7$ (Fig.~\ref{fig:DOS1}) even with the same degree of trigonal distortion of IrO$_6$ octahedra. It is clear from these results that the inter-site hopping, rather than the crystal field, plays a predominant role in the mixing of $J_{\mathrm{eff}} = 1/2$ and $J_{\mathrm{eff}} = 3/2$ states. In Fig.~\ref{fig:DOS2}(b), a clear splitting of $J_{\mathrm{eff}} = 3/2$ bands of about \SI{0.1}{\eV} is observed even for the undistorted Y model compound, which evidences that the effect of crystal field originating from neighboring cations is not negligible~\cite{Hozoi2014}. The splitting is enhanced by introducing the trigonal distortion observed experimentally, implying that the crystal field from both oxygen atoms and neighboring cations plays a significant role in the splitting of the $J_{\mathrm{eff}} = 3/2$ bands. Both In and Y model compounds with the experimentally observed trigonal distortion show a splitting of the $J_{\mathrm{eff}} = 3/2$ states of $\sim \SI{0.15}{\eV}$, which is smaller than that observed in the experiment ($\Delta{E} = \SI{0.4}{\eV}$). This may point to the involvement of other mechanisms, such as the molecular orbital formation due to the Ir-Ir and Ir-O hybridization, in the splitting of the $J_{\mathrm{eff}} = 3/2$ manifold.

The distinct electronic state of In$_2$Ir$_2$O$_7$ can be rationalized by the fact that the $t_{2g}$ bands for In$_2$(Ir$_{1/16}$Hf$_{15/16}$)$_2$O$_7$, as shown in Fig.~\ref{fig:DOS2}(a), acquire dispersion in contrast to Y$_2$(Ir$_{1/16}$Hf$_{15/16}$)$_2$O$_7$ (Fig. S7 in ~\cite{SupplementaryData}).  The hybridization of In 5$s$ and 5$p$ states with O 2$p$ states is not small due to the covalent character of In-O bond, which provide an extra hopping path between the remote Ir atoms in the model super-cell lattice. In In$_2$Ir$_2$O$_7$, the Ir-O-Ir bonds, rather than those via the In-O bonds, dominate the inter-site hopping. We argue that the effect of In-O hybridisation manifests itself differently: the presence of covalent In-O bonds reduces the Ir-O hybridisation and the inter-site hopping via Ir-O-Ir bonds. This explains the smallest bandwidth and the pronounced $J_{\mathrm{eff}} = 1/2$ character found in In$_2$Ir$_2$O$_7$ (Fig.~\ref{fig:DOS1}). The covalent In-O bond may also mask the effect of crystal field of In$^{3+}$ ion acting on Ir atom.

In summary, In$_2$Ir$_2$O$_7$ is a Mott insulator with nearly pure $J_{\mathrm{eff}} = 1/2$ character. The orbital states in pyrochlore iridates can be described as the interplay of three effects: trigonal crystal field, inter-site hopping, and A-site covalency. The crystal field comprises the contributions from the distortion of IrO$_6$ octahedra as well as Ir and A cations, and splits the $J_{{eff}} = 3/2$-derived states. This crystal field does not introduce substantial $J_{\mathrm{eff}} = 3/2$ character into the $J_{\mathrm{eff}} = 1/2$ band. The inter-site hopping instead plays a central role in mixing of the $J_{\mathrm{eff}}$ states. The covalent In-O bonds reduce the Ir-O hybridization and suppress the inter-site hopping of Ir $d$-electrons and thus the orbital mixing, resulting in the nearly pure $J_{\mathrm{eff}} = 1/2$ state. The sizable inter-site hopping in the other A$_2$Ir$_2$O$_7$ in contrast gives stronger mixing of $J_{\mathrm{eff}} = 1/2$ and $J_{\mathrm{eff}} = 3/2$ states than In$_2$Ir$_2$O$_7$. Therefore, the effects of inter-site orbital mixing and bond covalency, in addition to local structural distortion, need to be properly incorporated in order to understand their electronic structure. The importance of these factors would not be limited to pyrochlore iridates, and should be common in a wide variety of 5$d$ and 4$d$ transition-metal compounds with spatially extended $d$-orbitals. Those effects should be taken into account in designing materials hosting novel spin-orbit-entangled phases such as topological semimetal or quantum magnet.

\begin{acknowledgments}
We are grateful to A. V. Boris, T. Larkin, K. Rabinovich, R. K. Kremer and G. M. McNally for fruitful discussions. We thank E. Buchner, F. Falkenberg and K. Schunke for experimental support. The synchrotron radiation experiments were performed at the BL11XU of SPring-8 with the approval of the Japan Synchrotron Radiation Research Institute (JASRI) (Proposal No. 2016B3552). This work was partly supported by the Japan Society for the Promotion of Science (JSPS) KAKENHI (No. JP15H05852, JP15K21717 17H01140), the JSPS Core-to-core program "Solid-state chemistry for transition-metal oxides" and the Alexander von Humboldt foundation.
\end{acknowledgments}


\begin{thebibliography}{33}
\bibitem [{MH()}]{MH}%
	\BibitemOpen
	\href@noop {} {}\bibinfo {note} {The magnetization data included a very small
		ferromagnetic signal in all measured temperature range in addition to the one
		below \SI{55}{\kelvin}, most likely originating from tiny amount of unknown
		ferromagnetic impurity. The ferromagnetic moment was estimated to be $\sim$
		\SI{0.18} emu/mol from the magnetization curve at \SI{100}{\kelvin}. We
		subtracted this ferromagnetic component from the raw data by assuming that
		its temperature dependence is negligibly small}\BibitemShut {NoStop}%
\end{thebibliography}
\end{document}